\documentclass[12pt]{iopart}

\usepackage{iopams}
\usepackage{amstext}
\usepackage{braket}

\usepackage{todonotes}

\begin{document}
\title{Orders of chaoticity of unitaries}
\author{Adrian Ortega$^1$, Andrew B. Frigyik$^2$ and M\'aty\'as Koniorczyk$^1$}
\address{$^1$Department of Quantum Optics and Quantum Information, Institute for Solid State Physics and Optics, Wigner Research Centre for Physics, 1525 P.O. Box 49, Hungary}
\address{$^2$\'Obuda University, 1081 Budapest N\'epsz\'inh\'az st. 8, Hungary}
\ead{koniorczyk.matyas@wigner.hu}
\date{\today}

\begin{abstract}
  We introduce the concept of $K$-th order chaoticity of unitaries, and analyze it for the case of two-level quantum systems. This property is relevant in a certain quantum random number generation scheme. We show that no unitaries exist with an arbitrary order of chaoticity. 
\end{abstract}
\vspace{2pc}
\submitto{\PS}

\section{Introduction}

The quantification of the rate at which a signal source can produce information has always been one of the fundamental questions of information theory~\cite{Cover2005}. The celebrated result of Shannon states that a signal source that can be modeled with a sequence of independent identically distributed random variables (emitting
letters one after another) produces information that amounts to the entropy of the variable. If the output of the source is encoded in long sequences, this is the amount of bits per letter
minimally needed for 
encoding it, in an asymptotic sense. Entropy thus characterizes the lossless compressibility of the source's output.

If the variables in the sequence are not independent, instead of the entropy, the entropy rate  
becomes relevant. This quantity takes into account the correlations between the letters. 
In order to use the source in question as a random signal generator we want it to have the maximal possible entropy rate since it is the entropy rate that characterizes the amount of generated randomness. The 
idea of entropy rate can be 
generalized to the continuous setting, leading to the concept of Kolmogorov-Sinai entropy \cite{martin2011mathematical,walters2000introduction}.

Considering quantum dynamical systems, the generalization of the Kolmogorov-Sinai entropy is not obvious and can be approached from different points of view. The generalizations to non-commutative dynamical systems  by Connes, Narnhofer and Thirring~\cite{Connes1987} and 
the one by Alicki and Fannes~\cite{Alicki1994} are mathematically sound, but their operational meaning is less transparent than what we have in the classical case. In addition, they both vanish for finite dimensional quantum systems. The possible operational meaning highly depends on certain 
aspects of the given problem that has to be addressed: the freedom 
of choosing the dynamics, protocols, and measurements 
introduces a significant amount of ambiguity into the issues surrounding  
application and interpretation. S\l omczy\'nski and \. Zyczkowski~\cite{Slomczyski1994}, for instance, introduce the notion of `coherent states entropy' in order to study a certain aspect of quantum chaos. Quantum dynamical entropy, used in the sense of `amount of uncertainty in measurement results' has also been studied by many authors, e.g.~\cite{Srinivas1978,PhysRevA.46.6265,Crutchfield2008,PhysRevA.89.022338}. 

Especially since some of the quantum entropy definitions are only reasonable for systems with infinite dimensional Hilbert spaces, there are not many contributions 
concerning protocols involving simple finite dimensional quantum systems  in terms of quantum dynamical entropy. One of these is due to Alicki et al.~\cite{Alicki2004} who relate a partial entropy related to the Alicki-Fannes entropy but meaningful for finite dimensional systems to decoherence rate along with an illustration how it works on a particular simple model. 
S\l omczy\'nski and Szczepanek~\cite{Slomczynski2017} discuss a simple protocol involving qubits or qutrits. They consider an iterated dynamics scheme:  
an application of a unitary and subsequently a measurement, in each step. 
This setup can also be considered as a theoretical model of a quantum random number generator. They calculate a certain kind of dynamical entropy in this setting which, very 
obviously, 
is related to the performance of a unitary operator in such a random number generator scheme. This entropy rate is characteristic to unitary maps. 

A unitary that has the ability to implement a perfect random number generator --- i.e., if it can be used, along with a suitably chosen measurement, to produce sequences of independent uniformly distributed random variables with maximal entropy rate ---  is termed "chaotic unitary".
Somewhat surprisingly,
not all unitary qubit and qutrit operators are chaotic,
even though there is a significant manifold of suitable operators
including the most commonly considered ones.

Our present contribution aims to generalize the results of S\l omczy\'nski and Szczepanek to introduce a hierarchy of structural properties of unitaries: the notion of a unitary being "chaotic to the $K$th order". We will find that even if the random number generator is based on the repeated application of a qubit unitary and a subsequent measurement, the emerging hierarchy of randomness generation ability leads us to a
nontrivial structure of the set of unitaries. 

\section{Method}
\label{sec:method}

\subsection{PVM dynamical entropy}

Consider some iterated discrete-time dynamics of a quantum system. The
Hilbert space of the system is $\mathcal{H}$, which we assume to be finite dimensional, that is 
$\dim \mathcal{H} = d < \infty$. The initial state of the system is described by the (Hermitian, positive semidefinite, unit-trace) density operator
$\varrho^{(0)}$, and it evolves according to
\begin{equation}
  \label{eq:iterdyn}
  \varrho^{(k+1)} = \mathcal{E}\left[\varrho^{(k)}\right],
\end{equation}
where $\mathcal{E}$ is a completely positive trace preserving map. The overall goal is to describe the ``amount of randomness'' generated in such a process, or a related one eventually disturbed by measurements.

As was said before, the characterization of the randomness in
information-theoretic terms is ambiguous. One possible choice is the quantity
introduced by S{\l}omczy{\'{n}}ski and
{\.{Z}}yczkowski~\cite{Slomczyski1994, Slomczynski1995}, and it has 
been
calculated for unitary dynamics of 2 and 3 dimensional systems by
S{\l}omczy{\'{n}}ski and Szczepanek~\cite{Slomczynski2017}. This will
be our starting point. 

This approach assumes that a measurement is made after each step of the evolution.
Take a rank-1 POVM measurement $M$ with
outcomes $1\ldots k$ on a $d$-dimensional quantum system
which is characterized by the pure states
$\ket{\phi_j}\in \mathcal{H},\ j=1,\ldots, k$ so that
\begin{equation}
  \label{eq:meascompl}
  \sum_j \ket{\phi_j}\bra{\phi_j} = \frac{k}{d} \mathbb{I}.
\end{equation}
For $k=d$ we get a projector-valued measurement (PVM).
The probability of obtaining the
$j$-th measurement result if the system is in the state $\varrho$ is
\begin{equation}
  \label{eq:measprob}
  p_j = \frac{d}{k}\braket{\phi_j | \varrho | \phi_j}
\end{equation}
and the system is left in the state $\ket{\phi_j}$ after the
measurement. Hence, the model is such that the iterated dynamics described by
$\mathcal{E}$ is interrupted by the measurement after each evolution
step governed by $\mathcal{E}$. Clearly as the measurement modifies the system's state, the process will differ from the one defined in (\ref{eq:iterdyn}), yet this approach will finally lead to a quantity characteristic for $\mathcal{E}$.

The sequence of measurement outcomes
forms a stochastic process $\mathbf{X}=X_0,X_1, \ldots$,
to which the
approaches of information theory can be applied. In particular it is
possible to calculate its entropy rate
\begin{equation}
  \label{eq:entrate}
  H(U, M) = H(\mathbf{X})=\lim_{l\to \infty } \frac{\eta(X_1,\ldots X_l)}{l}
\end{equation}
describing the asymptotic minimum of the bits required to encode a
symbol of such a process when maximally compressed losslessly. This quantity depends on the unitary $U$ and the measurement $M$.  Here
$\eta$ is the Shannon-entropy function
\begin{equation}
    \eta(\mathbf{X})=- \sum_{\mathbf{x}} p(\mathbf{x})\log_2p(\mathbf{x}),
\end{equation}
where $\mathbf{x}$ runs through all possible values of $\mathbf{X}$, and $p(\mathbf{x})$ is the probability of obtaining $\mathbf{x}$.

In~\cite{Slomczynski2017} 
only unitary evolutions and PVM measurements were considered:
\begin{equation}
  \label{eq:unitary}
  \mathcal{E}(\varrho) = U \varrho U ^\dag,
\end{equation}
in which case we just obtain a classical Markov-chain with the
probability transition matrix
\begin{equation}
  \label{eq:ptrans}
  P_{i\to j}=|\braket{\phi_j|U|\phi_i}|^2.
\end{equation}
The PVM entropy rate is defined as
\begin{equation}
  \label{eq:PVMentrate}
  H(U) = \max_{M\in PVM}H(U, M) 
\end{equation}
(note that the measurements are restricted to PVMs) which, as derived in \cite{Slomczynski2017} on the basis of Eqs.~(\ref{eq:entrate}) and~(\ref{eq:ptrans}), calculates as
\begin{eqnarray}
  \label{eq:PVMentrate_caclc}
  H(U) & = \frac{1}{d}\max_{(\phi_j)_j \text{ ONB}}\sum_{j,l=1}^d \eta\left( | \braket{\phi_j|U|\phi_l}|^2\right)  \nonumber \\
    & = \frac{1}{d} \max_{V\in U(d)} \sum_{j,l=1}^d \eta\left( |(V^\dag U V)_{j,l}|^2 \right), 
\end{eqnarray}
where ONB stands for orthonormal basis, and $U(d) $ is the set of $d$-dimensional unitaries.

This quantity is calculated analytically in~\cite{Slomczynski2017} for
all unitaries in $d=2$ in closed form, and it is also studied in detail for  and $d=3$. For bigger systems it
can also be calculated via a numerical optimization over the unitary
group.
In particular, according to~\cite{Slomczynski2017}, for a given $2\times 2$ unitary, writing it in its eigenbasis as
\begin{equation}
  \label{eq:u2eig}
  U=
  \left(\matrix{%
  %\begin{pmatrix}
    \exp{i\phi} & 0 \cr
    0 & \exp{i\psi}
    % \end{pmatrix}
    }\right),
\end{equation}
with $\phi, \psi \in [0, 2\pi[$
and introducing
\begin{equation}
  \label{eq:theta}
  \theta = \min ( |\phi - \psi|, 2\pi - |\phi - \psi| ),
\end{equation}
we have
\begin{equation}
 \label{eq:Hqubit}
  H(U)=
   \cases{
    1, \quad \theta \geq \frac{\pi}{2} \\
    \eta\left(\cos^2\left(\frac{\theta}{2}\right)\right) +
    \eta\left(\sin^2\left(\frac{\theta}{2}\right)\right), \quad
       \theta \leq \frac{\pi}{2},
   }
\end{equation}
with $\eta(x) = -x \log_2 x$ for $x>0$ and $\eta(0)$=0.
Regarding chaoticity, Corollary 1 of~\cite{Slomczynski2017} implies that
\textit{$U$ is chaotic if and only if}
\begin{equation}
    |{\rm tr}(U)|\leq\sqrt{2},
    \label{eq:corollary},
\end{equation}    
or, equivalently,  $\theta\geq \pi/2$. 
We will use these facts as the basis of our considerations. 

\section{Results}

\subsection{$K$-sampled PVM dynamical entropies}

In general the process of measurement in each time step of an iterated unitary evolution proposed in~\cite{Slomczynski2017} can be considered as a first step towards a more general approach. 
Consider a modified protocol in which we skip every other measurement. In principle we use the square of the original unitary in this case. Pauli operators, for instance, are chaotic, but 
their square is the identity. So measuring in every other step generates no randomness at all.  This also means that if we have  
a random number generator based on Pauli operators and measurements  
then skipping some measurements will lead to correlations.

In the present contribution we deal with the modified protocol in which the measurement is performed
after each $K$th
iteration only.
We define the \emph{$K$th order dynamical entropy} as the entropy of
$U^K$:
\begin{equation}
  \label{eq:KPVM}
  H_K(U)=H(U^K).
\end{equation}
(The definition could be extended to more general completely positive trace preserving maps.)
It is reasonable to ask whether there are evolutions for which $H_K$ is nonzero (or
 even maximal) for some or all values of $K$? In what follows we will study this question for the case of 2-dimensional unitaries. 
 
 We will call a unitary $U$ \emph{chaotic
 to the $K$th order}, if $H_K(U)$ is maximal. For instance, the Pauli operators are chaotic to all odd orders, but their even order dynamical entropies are all zero 
 for they are idempotent. 
 We will call a unitary $U$ \emph{chaotic to an arbitrary order}, if $H_K(U)$ is maximal for any $K$. Such a unitary could be very useful in random generation
 for it does not 
 create the need to carry out the measurement in every iterative step in order to obtain a proper generator. 
 
 The idempotent nature of a matrix is clearly a stronger property than nonchaoticity. Hence, the notion of an idempotent and a non-idempotent 
 matrix will be useful for our study. For a unitary $U$, if $U^K=\mathbb{I}$ for some $K>0$ integer, and $K$ is the lowest integer for which this property holds, then we say that $U$ is idempotent of order $K$. (If $U$ is idempotent to the order $K$ then trivially $U^{N\,K}=\mathbb{I}$ for arbitrary $N\in \mathbb{Z}$.) We say that a non-idempotent matrix is a matrix such that for all $K$, $U^K\neq \mathbb{I}$.

As dynamical entropy and thus the notion of chaoticity is phase invariant, instead of idempotency, the notion of "phase idempotency", that is, $U^K=\exp{(i\varphi)}\mathbb{I}$ for some $K$ and $\varphi$, could also be considered instead of idempotency. It is also a stronger property than nonchaoticity to the $K$-th order, also excluding the latter. Moreover as it is a more general notion of idempotency, it could reveal more non-chaotic unitaries. We leave its consideration to future studies and address idempotency in what follows. In case of $SU(2)$ matrices, for instance, it is easy to see phase idempotency means $U^K=\pm\mathbb{I}$; thus we will not consider the negative sign.

The $K$th order dynamical entropy and the corresponding chaoticity can be addressed on the basis of the results in 
\cite{Slomczynski2017}: 
In order to calculate $H_K(U)$, in Eq.~(\ref{eq:theta}) we have to replace $\phi$
and $\psi$ by $K\phi$ and $K\psi$, restricted back to $[0, 2\pi[$. This restriction is not trivial because of the form of Eq.~(\ref{eq:theta}). To decide whether the unitary is chaotic to the $K$-th order, according to~(\ref{eq:corollary}), it is sufficient to check if
\begin{equation}
    |{\rm tr}(U^K)|\leq\sqrt{2}.
    \label{eq:corollaryK}
\end{equation}
Before addressing chaoticity to the $K$th order as well as
idempotency, let us first restrict ourselves to the group of unimodular
unitaries, and discuss $K=1$ in that case.

\subsection{Eigenphase distribution of chaotic unitaries in the group $SU(2)$
}
\label{sec:epdist}

In the rest of our considerations we 
restrict our discussion 
to the group
$SU(2)$. We will not lose generality as any unitary has a
unimodular counterpart that leads to the same physical behavior up to
an irrelevant global phase.
Let us also remark that the elements of this
subgroup of unitaries are readily  
implementable on some real
quantum computers, such as the IBM, IonQ and Rigetti
platforms~\cite{IBMqcs,ionqqc,rigettiqc}.

The restriction to $SU(2)$ leads to a significant simplification: for the two phases characterizing the unitary in the form as in~Eq.(\ref{eq:u2eig}),
\begin{equation}
    \label{eq:su2ph}
    \phi + \psi = 2\pi m
\end{equation}
holds, where $m$ is a natural number. Hence, we have a single phase parameter instead of two, and an integer playing a simpler role.From Eq.~(\ref{eq:su2ph}) it follows that $\phi = 2\pi m -\psi$;  
substituting it into 
Eq.~(\ref{eq:u2eig}), the condition of chaoticity in Eq.~(\ref{eq:corollaryK}) 
reads:
\begin{equation}
\label{eq:psicond}
    |\cos\psi| \leq 2^{-1/2}.
\end{equation}

 Note that the use of the single parameter $\psi$ restricts the considered set of unitaries to a subgroup of $SU(2)$; a great circle in $S^3$. Every other $SU(2)$ element is  conjugate to one of the elements of this subgroup~\cite{faraut2008analysis}. Hence, using this parameter is consistent with the Haar measure of the group, justifying our discussion of distributions or probabilities.

This last inequality is the defining condition for $SU(2)$ matrices to be chaotic, i.e. they are chaotic if $\psi\in[\pi/4,3\pi/4]\cup[5\pi/4,7\pi/4]$. The length of the intervals sum up to $\pi$, which means that half of the $SU(2)$ matrices defined through $\psi$ are chaotic and half of them are non-chaotic. In other words, if we draw 
uniformly a $\psi$ phase and construct the corresponding
$SU(2)$, with probability $1/2$ it will be chaotic. 
Thus, the probability of obtaining $k$ instances of chaotic matrices (of the first order) out of $N$ trials 
has a binomial distribution (or a normal distribution in the limit as $N$ goes to infinity).

\subsection{Idempotency of arbitrary order and chaoticity of order $K$}

Some of the prominent examples of chaotic unitaries found in~\cite{Slomczynski2017}, such as the Pauli operators, are idempotent of order two, i.e. 
their square is the identity operator, hence they cannot be chaotic to the second order. Idempotency of order $n$ excludes chaoticity to the order of $n$. Therefore, let us address the question of idempotency
of  
order $n$ where $n$ is arbitrary.
(This implies, by definition, that $n$ is  the lowest 
such value for which $U^n = \mathbb{I}$.)  
Based on Eq.~(\ref{eq:u2eig}), we can 
express a certain class of unitaries as
\begin{equation}
    \label{eq:Ufrac}
    U \propto \left( \matrix{e^{im_1\pi/p_1} & 0\cr 0 &  e^{im_2\pi/p_2}}\right),
\end{equation}
where $m_i,p_i$ ($i=1,2$) are integers and 
$m_i$ and $p_i$ are relative primes. With these conditions it is easy to produce examples with arbitrary order of idempotency, for instance,
\begin{equation}
    D^{(4)}=e^{i\pi/4}\mathrm{Diag}(e^{i\pi/4},e^{i5\pi/4})
\end{equation}
is idempotent of order 4 and 
\begin{equation}
    D^{(8)}=e^{i23\pi/32}\mathrm{Diag}(e^{i\pi/32},e^{i17\pi/32})
\end{equation}
has idempotency of order 8. Furthermore, in the first case $\theta=\pi$ while $\theta = \pi/2$ in the second case, and thus the latter has the same degree of chaoticity as $\sqrt{\sigma_x}$-gate (period 4). Note that one can always enforce an 
arbitrarily (or indefinitely) long period,  
provided that the $p_i$'s are very large different primes with $m_i \ll p_i$ (preferably $m_i = 1$ for $i=1,2$). Indeed, the order of idempotency of the operator in this case is $n={\rm lcm}(p_1,p_2)$, 
where ${\rm lcm}$ is the least common multiple function.

\begin{figure}[htb]
  \centering
  \includegraphics[scale=1.5]{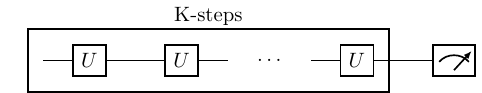}
  \caption{Chaoticity of order $K$: Using condition Eq.~(\ref{eq:condK}) we can construct a unitary $U$ such that its $K$-th power is chaotic: the measurements are performed in each $K$th step.}
  \label{fig:circ1}
\end{figure}
Idempotency to the order $n$ implies that $H_K(U)=0$ for all $K$s that are multiples of $n$. However, we are in search of chaotic unitaries, hence, we also require that in a certain step $K$, $U^K$ ($\neq \mathbb{I}$) 
should be a chaotic unitary, c.f. Fig.~\ref{fig:circ1}. One way to achieve this is to choose a rational eigenphase of $U$ such that $m_2=1$, $p_2$ is a prime and $K$ does not contain $p_2$ as a factor 
while
\begin{equation}
    \label{eq:condK}
    \left|\cos\left(\frac{\pi K}{p_2}\right)\right|\leq\frac{1}{\sqrt{2}}.
\end{equation}
The phase $\phi$ is determined by Eq.~(\ref{eq:su2ph}); if we choose it as in Eq.~(\ref{eq:Ufrac}), then $m_1/p_1 = 2m - 1/p_2$. 
This is the condition to construct a unitary that is chaotic of a given order $K$.
As an example, choose $K=5$ and $p_2=2$. Then $\psi = \pi/2$, $\phi = 3\pi/2$ and
\begin{equation}
    U^5 = \left( \matrix{e^{i15\pi/2} & 0\cr 0 &  e^{i5\pi/2}}\right)
\end{equation}
is chaotic with $\theta=\pi$.

Note that unitaries with rational frequencies as in Eq.~(\ref{eq:Ufrac}) are the exceptions more than the rules in the group $SU(2)$. Following the discussion of the previous section, rational frequencies $\psi = m_2/p_2$ form a subset of (Lebesgue) measure zero and thus in a random trial we are always going to draw a $SU(2)$-matrix without a rational frequency. Hence in principle this larger set of $SU(2)$ matrices is non-idempotent, but it remains to be determined if a matrix in it is chaotic or not %at
for a certain $K$. We shall investigate in the following this %largest
larger subset of $SU(2)$ matrices.

\subsection{Chaoticity to arbitrary order}
\label{sec:arbord}

\begin{figure}[htb]
  \centering
  \includegraphics[scale=1.5]{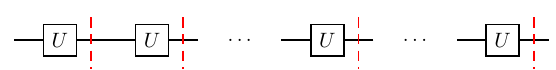}
  \caption{Chaoticity to arbitrary order. The dashed lines indicate potential measurements; any of them can be carried out or skipped. The result sequence is composed of the outcomes of the subsequent measurements that were actually carried out.
  }
  \label{fig:circ2}
\end{figure}

Recall that unitaries chaotic to all orders are chaotic to the $K$-th order for any $K$, and thus they have to be non-idempotent.
Figure~\ref{fig:circ2} illustrates a circuit with such a $U$. The dashed lines illustrate that we can measure at any step of the circuit and the sequence of the so obtained measurement outcomes will have a maximal entropy. We shall discuss in the following if such chaotic unitaries exist.

First let us check the condition of non-idempotency. In order to see if one can obtain a non-idempotent matrix 
in the group $SU(2)$, we set $\phi=x\pi$ and $\psi=y\pi$ in Eq.~(\ref{eq:su2ph}). Thus the real numbers $x,y$ must fulfill
\begin{equation}
\label{eq:su2cond}
    x + y = 2m.
\end{equation}
If both $x$ and $y$ are rational numbers, then there will be a power such that if we raise  
the matrix to that power we get  
the identity. Hence, in order for an element of $SU(2)$ 
to be non-idempotent at least one of the members of the pair $(x,y)$ has to be irrational. Meanwhile the condition in Eq.~(\ref{eq:su2cond}) also holds, hence, if one of them is irrational, so is the other, as their sum is an (even) integer. 

From the arguments given so far, we can come to an important consequence: \emph{there exist no element
that
is chaotic to arbitrary order within the group $SU(2)$}. If such a matrix existed, it would have to be non-idempotent.  
If $x,y$ in Eq.~(\ref{eq:su2cond}) are irrationals, then by the Kronecker's approximation theorem~\cite{apostolmodular}, $\exp(iK\pi y)$ will fill densely the unitary circle. This means that for some $K$'s, the condition Eq.~(\ref{eq:psicond}) will be violated and, if we do the measurement at this step, the matrix will be non-chaotic. In fact there will be infinitely many $K$-s for which $H(U^K)$ will be arbitrarily close to zero.

The question arises how to generate particular non-idempotent matrices that are candidates in search for matrices chaotic for given orders. Examples of phases with the property described in Eq.~(\ref{eq:su2cond}) can be drawn straightforwardly for real quantities such as $x = q_1 + \sqrt{a}$ and $x = q_2 - \sqrt{a}$, where $q_1,q_2\in\mathbb{Q}$, $q_1 + q_2 = 2m$ and $\sqrt{a}$ is not an integer. 
In the following we use these type of numbers to show an interesting connection between our problem
and the field of algebraic number theory.
We describe two types of series of non-idempotent $SU(2)$ matrices. In general, in one of them the matrices converge to the identity as the parameter grows, while the other ``traverses" all $SU(2)$.

Let us give an example  
from the set 
consisting of  
the first kind of matrices, as a motivation. Assume that
$x^{1/t}=(1+\sqrt{5})/2$ and $y^{1/t}=(1-\sqrt{5})/2$, where $t$ is a positive integer. This is the
golden ratio and its conjugate, with $t$ a positive integer such that
\begin{equation}
    (x^{1/t})^t + (y^{1/t})^t = 2m.
\end{equation}
The numbers defined above are the celebrated Lucas numbers~\cite{Andrewsbook1994}
\begin{equation}
L_t = \lambda_+^t + \lambda_-^t,\label{eq:Lucasdef}
\end{equation}
with $\lambda_\pm = (1\pm\sqrt{5})/2$, where we are 
considering only those $t$ values for which $L_t$ 
is even. The first non trivial even Lucas number is $L_3 = 4$, which implies $\phi = (\lambda_+)^3\pi \approx 0.7416$ and $\psi = (\lambda_-)^3\pi \approx 5.5415$ but
\begin{equation}
    |e^{i\phi} + e^{i\psi}|\approx 1.4747,
\end{equation}
and this is bigger than $\sqrt{2}$, which is the lower bound for a
matrix to be chaotic. Yet, since 1.4747 is close to $\sqrt{2}$, we may
consider the unitary with above eigenphases as close to chaotic: the actual value of the dynamical entropy calculated using~(\ref{eq:Hqubit}) is $0.9944$, which is indeed close to the maximum value of $1$. Note that
for large enough $t$, $\lambda_-^t \approx 0$, $\lambda_+^t \approx 2m$
from Eq.~(\ref{eq:su2cond}) and
$|e^{i\phi} + e^{i\psi}|\approx |e^{i\phi} +1 | = 2$ will always yield
a non-chaotic matrix.

After this illustration let us generalize our
selection in order to show how to  
construct in a more general way series of non-idempotent $SU(2)$ elements. The 
process results in two series, one of which converges to the identity, and  
the other traverses 
the whole $SU(2)$.
The numbers $\lambda_\pm$ that enter 
the definition of the Lucas numbers are solutions of the quadratic equation $x^2 -x - 1=0$. We can generalize easily this result in order to obtain a whole class of non-idempotent $SU(2)$ matrices. Let $\alpha,\beta$ be solutions of the quadratic equation
\begin{equation}
    x^2 +a x + b = 0,
\end{equation}
where $a,b\in\mathbb{Z}-\lbrace 0\rbrace$. Obviously its solutions are $\alpha = (-a  +\sqrt{a^2 - 4b})/2$, $\beta = (-a  -\sqrt{a^2 - 4b})/2$ (if the discriminant $D = a^2-4b$ is square-free (its prime factorization is a product of distinct primes where each factor is to the power of, at most, one) then $\alpha,\beta$ belong to the quadratic field $\mathbb{Q}(\sqrt{D})$~\cite{Nivenbook1991}). Note that
\begin{equation}
    \alpha^t + \beta^t = m
\end{equation}
is always an integer. This last equation can be proven by induction taking into account that $\alpha+\beta = -a$ and $\alpha\beta = b$ are integers and thus, if $\alpha^t + \beta^t$ is an integer,
\begin{equation}
\label{eq:indst}
    \alpha^{t+1} + \beta^{t+1} = (\alpha^t + \beta^t)(\alpha + \beta) - \alpha\beta (\alpha^{t-1} + \beta^{t-1})
\end{equation}
is also an integer. In Eq.~(\ref{eq:su2cond}), set $x=\alpha^t$ and $y=\beta^t$, i.e. we want to have
\begin{equation}
    \label{eq:ev}
    \alpha^t + \beta^t = {\rm even}.
\end{equation}
And this does hold: to prove it one can use induction again and notice that if $\alpha + \beta$ is even, using Eq.~(\ref{eq:indst}) yields that Eq.~(\ref{eq:ev}) is even, irrespective of the value of $\alpha\beta$ (c.f. $L_t$ is even for \textit{some} values of $t$). For simplicity we consider from here onward
the case when $a,b\leq0$. Furthermore, in order to have non-idempotency we choose $a,b$ such that $\sqrt{a^2 + 4|b|}$ is not an integer.
A non-idempotent  $U$
is attained by choosing the algebraic numbers $x=\alpha^t$ and $y=\beta^t$ constructed as above. Here, $t$ is a parameter for which one can choose $|\cos\psi|\leq 2^{-1/2}$ to obtain a chaotic unitary, c.f. Sec.~\ref{sec:epdist}. 

On to the other type of series:  
those that traverse  
the whole $SU(2)$, if we want to avoid $\beta^t \rightarrow 0$ in the limit of 
large $t$, we need to require that $\beta < -1$. For an example, choose $|a|=2$, $|b|=101$ (both primes) and $t=8$; this yields $|\cos \psi|\approx 0.387 < 2^{-1/2}$ and thus $U$ is an $SU(2)$ non-idempotent matrix with eigenphases $\phi = \pi(2 + \sqrt{2^2 + 4\times 101})^8/2^8$, $\psi = \pi(2 - \sqrt{2^2 + 4\times 101})^8/2^8$. 

\section{Discussion}
\label{sec:discussion}

In this section first, we focus on two immediate outcomes of the results presented in the previous section. 
Next we address the question 
of how a simple model of noise affects the chaoticity of a generic $SU(2)$ matrix.

A realistic implementation of a chaotic $SU(2)$ matrix will always be subject to noise coming from distinct sources. As of today, the sources of noise and errors in devices of the Noisy Intermediate-Scale Quantum (NISQ) era are complex and diverse~\cite{SalonikArxiv2019,AspuruRMP2022}.
In this section, we would like to understand what 
the impact of a small amount of noise 
on a chaotic $SU(2)$ matrix is. 

The following is a simple noise model: 
take a uniformly distributed random phase $\lambda$ 
from the interval $[-\epsilon\pi,\epsilon\pi]$ where $\epsilon$ is small and positive. Using Eq.~(\ref{eq:su2ph}) we modify the phases 
to
\begin{equation}
    \tilde{\phi} + \tilde{\psi} = (\phi+\lambda) + (\psi-\lambda).
\end{equation}
With this choice we still get an $SU(2)$ matrix, albeit with different eigenphases that we can modify at each time step. Each time we draw a random number $\lambda$ from the interval $[-\epsilon\pi,\epsilon\pi]$ we will get an irrational value (again the rationals in these intervals have measure zero). Thus the modified eigephases $\tilde{\phi}$ and $\tilde{\psi}$ will always be irrational in any step of the protocol. 
Aside from the non-idempotency added by $\lambda$, we also note from Eq.~(\ref{eq:psicond}) that if $|\cos\psi|$ is close to the value of $2^{-1/2}$ and $\epsilon$ is ``large enough", then in some trials we can alternate between chaotic and non-chaotic unitaries. Note 
that because of the way we defined the dynamics here, any multiplicative noise that rotates the unitary in question is irrelevant.
In other words, $U$ is defined up to an arbitrary (noisy) unitary transformation.

\section{Conclusion}

We have introduced the notion of chaoticity of a unitary to the $K$th order. We have discussed it in case of two-level quantum systems.
We have also studied idempotency which 
plays an important role when it comes to chaoticity. 

Our most important conclusion is that no unitaries exist that are chaotic to an arbitrary order. From the point of view of random number generation with the studied scheme, the situation is somewhat analogous to the classical Marsaglia-type pseudo random number generators~\cite{Marsaglia2003-wd}: they can have extremely long periods but they are periodic. 

A possible further generalization of this work could be the study of the entropy rate of a process arising from the same iterated unitary evolution but making measurements in unevenly or even randomly distributed discrete time instants only. For instance, one could investigate the situation when the measurement is made after \emph{at most} $K$ steps. This could be a model of a random generator in which the measurement is eventually skipped, e.g. due to failure, in a few consecutive steps. This may be a potential direction of future research.

In a future work, we would like to study the relationship between the chaotic unitaries defined here compared to the unitaries defined in Random Matrix Theory~\cite{GuhrPR1998}. It would be interesting to see if extra symmetries, such as the one present in the Circular Unitary and Orthogonal Ensembles, affect in any way the chaoticity presented here.

\section*{Acknowledgements}

This research was supported by the National Research, Development, and
Innovation Office of Hungary under project numbers K133882 and K124351
and the Ministry of Culture and Innovation and the National Research, Development and Innovation Office within the Quantum Information National Laboratory of Hungary (Grant No. 2022-2.1.1-NL-2022-00004). The idea of the presented research raised while visiting and discussing with Igor Jex in Prague.

$\quad$

%\bibliographystyle{iopart-num}
%\bibliography{qudyn}
\providecommand{\newblock}{}

\end{document}